   \documentclass[prl,aps,twocolumn,showpacs]{revtex4}
\usepackage[dvips]{graphicx}
\usepackage{dcolumn}
\newcommand{\be}{\begin{equation}}
\newcommand{\ee}{\end{equation}}
\newcommand{\bea}{\begin{eqnarray}}
\newcommand{\eea}{\end{eqnarray}}
\newcommand{\nn}{\nonumber}

\begin{document}

\title{Local Geometry of the Fermi Surface and the Cyclotron Resonance in Metals in a Normal Magnetic Field}

\author{Natalya A. Zimbovskaya$^{1,2}$ and Godfrey Gumbs$^3$}

\affiliation{$^1$Department of Physics and Astronomy,
St. Cloud State University,
720 Fourth Avenue South, St. Cloud, MN 56301; $^2$Department of Physics and Electronics, University of Puerto Rico at Humacao, Humacao, PR 00791; \\
$^3$Department of Physics and Astronomy
Hunter College of the City University of New York, 695 Park Avenue, New York, NY 10021 }

 \begin{abstract} 
 In this paper we present a detailed theoretical analysis of the cyclotron resonance in metals in the magnetic field directed along a normal to the surface of a sample. We show that this resonance occurs due to local geometry of the Fermi surface of a metal. When the Fermi surface (FS) includes segments where its curvature turns zero, this could give rise to resonance features in the frequency/magnetic field dependence of the surface impedance or its derivative with respect to the field. Otherwise the resonance is scarcely detectable unlike the well-known cyclotron resonance in a parallel magnetic field. The proposed theory agrees with experimental results concerning both convenient and organic metals.
 \end{abstract}

\pacs{71.18.+y, 71.20-b, 72.55+s}
\date{today} 
\maketitle

\section{I. Introduction }
\label{sec1}

High-frequency properties of metals in the presence of an
external magnetic field were repeatedly analyzed in detail. It is well known that a periodical motion of conduction electrons in the magnetic field can cause a resonance with the electric field of an incident electromagnetic wave. This resonance occurs when the cyclotron frequency of electrons $ \Omega $ coincides with the frequency of the electromagnetic field $ \omega. $ There are two geometries providing the resonance features in the surface impedance to be displayed. At first, strong resonance arises when the magnetic field is directed nearly in parallel with the surface of a metal. The conduction electrons spiral around the magnetic field and at each revolution some of them return to the skin layer where they gain energy from the electromagnetic wave. Another opportunity for the resonance appears when the magnetic field is perpendicular to the surface. In this case some conduction electrons remain within the skin layer for a long time absorbing the energy. As a rule, the cyclotron resonance in a normal magnetic field is not manifested in good metals. The usual explanation
is that the skin layer in metals is very thin at high frequencies. Therefore the percentage of conduction electrons moving within the skin layer in parallel with the metal surface is too small to provide a distinquishable resonance feature at $ \omega = \Omega $ in either the surface impedance of the metal or the absorbed power. This
was supported with thorough calculations carried out assuming the FS of a metal to be a smooth surface of a nonzero curvature (see e.g \cite{one}).
Nevertheless, the cyclotron resonance in the normal magnetic field was observed in potassium \cite{two}, cadmium and zinc \cite{three,four}. No detailed theory has yet been given for these old experiments.
Qualitative explanation of the experiments on potassium was offered basing on the assumption that the FS of this metal includes some cylindrical segments \cite{five} (this assumption agrees with the results concerning the FS shape in potassium which follow from the charge density wave theory \cite{six,seven,eight}, and was confirmed in experiments of \cite{nine,ten}).
Here, we adopt and further develop this approach.

The purpose of the present work is to show that the cyclotron resonance in the normal magnetic field could appear in metals when their Fermi surfaces (FS) possess some special local geometrical characteristics such as local flattenings, nearly cylindrical segments, fracture lines and some others. The point of the present analysis is the effect of the FS local geometry on the frequency dependence of the surface impedance. We prove that when the FS of a metal has lines of zero curvature they affect the anomalous skin effect characteristics. As a result, the cyclotron resonance in the normal magnetic field could be manifested in conventional metals in the same way as in layered organic metals. Cyclotron resonance in the normal field was observed in the experiments \cite{eleven,twelve} on the quasi-two-dimensional organic metals and there are grounds to believe that the effect sensibly depends on the local geometry of the FSs of these substances \cite{thirteen}. We also analyze possible manifestations of local flattening on the FS in the surface impedance. This enables us to semiquantitatively describe resonance features resembling those observed in cadmium and zinc \cite{three,four}. The outline of the remainder of this paper is as follows. In Section II, we analyze the effects of the FS local geometry on the electric conductivity tensor. In Section III, the surface impedance is calculated, and it is shown that local anomalies of the FS curvature could give rise to various resonance features in the frequency dependence of the impedance near $ \omega = \Omega.$ Finally, we discuss our results in Section IV.

\section{II. Electron Conductivity}

In further analysis we consider a semi-infinite metal which fills the half-space $ z < 0. $ We assume that the external magnetic field $ \bf B $ is directed along the normal to the surface of metal. The response of the conduction electrons to an electromagnetic disturbance of the frequency $ \omega $ and the wave vector $ \bf q $ (the latter is supposed to be parallel to "z" axis) could be expressed in terms of the electron conductivity components.

For a metal with a closed Fermi surface the latter could be written as follows \cite{fourteen}:
      \bea 
  &&\sigma_{\alpha \beta} = \frac{2 i e^2}{(2 \pi \hbar)^3} \sum \limits_n \int_0^{2 \pi} d \psi \int d p_z  \\ \nn \\ 
\!\! \! &\times& \!\!
 \int_{- \infty}^0 d \eta e^\eta
\frac{m_\perp(p_z) v_a(p_z, \psi) v_{n \beta}(p_z) \exp (- i n \psi)}{\omega + i/\tau - n \Omega (p_z) - q v_z
\big (p_z, \psi + \delta \psi (\eta)\big )}.\nn
            \eea
Here, $ v_{\alpha,\beta } $ are the electron velocity components; $ p_z $ is the longitudinal component of electron quasimomentum; $ m_\perp (p_z) $ is the electron cyclotron mass, and $ \tau $ is the electrons scattering time. The angular variable $\psi = \Omega t $ is related to the time $ t $ of electron motion along the cyclotron
path. For a metal with a complex many-sheet FS
(for example, cadmium or zinc), integration with respect to $ p_z $ in Eq.(1) is carried out within the
limits determined by the form of each sheet, and summation is carried out over all sheets of the Fermi surface. To analyze the cyclotron resonance we need an appropriate asymptotic expression for the conductivity components at large $ q $ when $ u = \omega / q v_F $ is small compared to unity $ ( v_F $ is the Fermi velocity). At $ u << 1 $ the correction $ \delta \psi (\eta) $ is nearly proportional to $ \eta $ and we can write the
relation: $ \delta \psi (\eta) \approx - i \eta \Omega /q v_z(p_z, \psi). $ To proceed we expand the last term in the denominator of Eq.(1) in a series in $ \delta \psi, $ and keep the first two terms:
  \bea 
qv_z \big (p_z, \psi + \delta \psi (\eta) \big ) & \approx & q v_z (p_z, \psi) - i \eta \Omega \frac{\big [\partial v_z (p_z, \psi ) \big / \partial \psi \big ]}{v_z (p_z, \psi)}   
 \nn \\ \nn \\
 & \equiv & qv_z (p_z, \psi) - i \eta \Delta (p_z, \psi).
            \eea

Then we change variables in Eq.(1) and pass to integration over velocities space. To carry out the integration we divide the FS in segments supposing that electron quasimomentum components are single-valued functions of the velocity magnitude $ v $ and angles $ \theta, \varphi $ determining the position of the velocity vector in spherical system of coordinates over each segment. This change of
integration variables gives the result:
  \bea
   &&
\sigma_{\alpha \beta} = \frac{2 i e^2}{(2 \pi \hbar)^3 q} \sum \limits_n \int_{-\infty}^0 d \eta
\sum \limits_j \int d \theta \int d \varphi
\frac{\sin \theta}{ |K_j(\theta, \varphi) |}    
     \nn \\   \\  
   &\times&
\frac{n_{\alpha j}(\theta, \varphi) V_{n \beta}^j
(\theta,\varphi)e^\eta}{\big |\omega - n \Omega_j (\theta,\varphi) + i \Delta (\theta, \varphi)
\eta + i/\tau | \big / q v_j(\theta, \varphi) -
\cos(\theta)},  \nn
            \eea
where $ K(\theta, \varphi) $ is the FS curvature,
  \bea 
V_{n \beta}^j (\theta, \varphi) & = &\frac{v_{n \beta j} (\theta, \varphi) \exp(- i n \psi)}
{v_j(\theta, \varphi) },
         \nn \\
n_{\alpha j}(\theta, \varphi) & = & \frac{v_{\alpha j}
(\theta,\varphi)}{v_j(\theta, \varphi)}.
       \nn \eea

The summation over $"j"$ is carried out over all FS segments. When $ u << 1 $ the main contributions to the conductivity components (3) comes from narrow effective strips on the FS where $ v_z = 0 \ (\theta = \pi /2). $
The principal terms in the expansion of the electrical conductivity tensor components in powers of $\omega /q v_F $ are independent of the magnetic field. For $ \sigma_{xx}^0, $ we obtain the familiar relation:   
   \be 
\sigma_{xx}^0 (q) = \frac{e^2}{4 \pi^2 \hbar^3 q} \sum \limits_j
\int \frac{\cos^2 \varphi d \varphi}{|K_j (\pi /2, \varphi)|} \equiv
\frac{e^2}{4 \pi \hbar^3 q} p_0^2.
            \ee
Again, summation over $ j $ is carried out over the effective strips on the FS, and $ K_j (\pi /2, \varphi ) $ is the Gaussian curvature at the corresponding
points on the effective $ j $-th strip. The principal term in the asymptotic expression for
the component $ \sigma_{yy} $ is described by a similar formula, where the numerator in the integrand in each term is replaced by $ \sin^2 \varphi. $ For a metal with everywhere finite nonzero curvature of the FS the first correction to the principal term in the expansion for $ \sigma_{xx} $ has the form
   \be 
\sigma_{xx}^{(1)}(\omega, q) = i \sigma_{xx}^0 (q) \frac{\omega}{q} \sum \limits_n \bigg <\frac{\chi_n}{v_0} \bigg >.
           \ee  
  Here
      \bea 
 &&\bigg <\frac{\chi_n}{v_0} \bigg >  = \frac{2}{\pi^2 p_0^2} 
\sum \limits_j \int_{-\infty}^0 e^\eta d \eta \int d \theta \int d \varphi
\frac{\cos \varphi \sin \theta}{\cos^2 \theta}      
  \nn \\ \nn \\  & \times &
\left [ \frac{V_{nx}^j (\pi /2, \varphi) \chi_{nj} (\pi /2,\varphi, \eta)}
{\big |K_j(\pi /2,\varphi)| v_j(\pi /2, \varphi)}
\left (1 + \frac{\cos^2 \theta}{\cos^2 \theta_j} \right )
        \right. 
 \nn \\ \nn \\ & - & \left. \frac{V_{nx}^j (\theta, \varphi) \chi_{nj} (\theta, \varphi, \eta)}
{\big |K_j (\theta,\varphi)\big | v_j(\theta, \varphi) } \sin \theta \right ]                -
\frac{2}{\pi^2 p_0^2} \sum \limits_{r \neq j} \int_{-\infty}^0 e^\eta d \eta         
  \nn \\ \nn \\  & \times & 
 \int d \theta \int d \varphi \frac{\cos \varphi \sin^2 \theta V_{nx}^r (\theta, \varphi) \chi_{n r}
(\theta, \varphi, \eta)}{\big |K_r (\theta, \varphi) \big | v_r(\theta, \varphi) }, 
            \eea
   where
     $$
\chi_{nr}(\theta, \varphi, \eta) = 1 - \frac{n \Omega_r (\theta, \varphi)}{\omega} + \frac{i \Delta_r(\theta, \varphi) \eta}{\omega} +
\frac{i}{\omega \tau}.
       $$

The last term in (6) takes into account contributions from those parts of the FS which do not contain effective lines. Integration with respect to $ \theta $ and $ \varphi $ in each term of the sums over $ r $ and $ j $ is carried out within the limits set by the shape and size of the corresponding segment of the Fermi surface. It follows from (6) that the quantity $ v_0 $ in the expression (5) has the dimensions
of velocity, while its value for a one-sheet FS is of the order of Fermi velocity of electrons. The first correction to the main approximation of the electrical conductivity component $ \sigma_{yy} $ is described by a similar expression.
 
  In general, we can present the conductivity in the form:
   \be 
\sigma_{xx} (\omega,q) = \sigma_0 (q) (1 + \Lambda_{1xx} u+ \Lambda_{2xx} u^2 + \dots).
   \ee
Here, the coefficients $ \Lambda_{nxx} $ are continuous functions of $\omega $ and $ \Omega. $ For instance, the expression for the coefficient $ \Lambda_{1xx} $ could be found from Eq.(6). The conductivity approximation based on the expansion similar to Eq.(7) was employed in the earlier paper \cite{thirteen} concerning organic metals. However, it is appropriate in general case provided that the FS curvature is finite and takes on nonzero values over the whole effective part of the FS.

Now, we start to analyze the effect of the FS local geometry on the electron conductivity. First we assume that the FS curvature turns zero at a whole effective line. We suppose that the curvature anomaly is attributed to the radius of curvature which is directed perpendicularly to the effective line. This means that the FS includes a nearly cylindrical segment whose curvature could be approximated as follows:
   \be 
K(\theta,\varphi) \sim (\cos\theta)^{-\beta} W (\theta,\varphi)
  \ee
where $ \beta < 0, $ and the function $ W (\theta,\varphi) $ accepts nonzero values.

The conductivity may be sensibly affected due the presence of the nearly cylindrical belts on the FS.The reason is that the FS curvature at an effective cross-section is closely related to a width of the associated strip, that is, to the number of effective electrons which belong to the strip. The number of conduction electrons belonging to nearly cylindrical strips $(\beta < 0)$
could be significantly greater than for "normal" effective strips, and this enhances the contribution from such strips to the conductivity. As a result a special term emerges in the asymptotic expression for the conductivity at large $q. $

To avoid tedious calculations we employ a model of axially symmetric FS in computation of this term. Assuming that the magnetic field is directed along the symmetry axis, both velocity magnitude and FS curvature do not depend on the angle $ \varphi. $ This enables us to carry out integration with respect to $ \varphi $ and $ \eta $ in the general expression (3). The conductivity tensor is diagonalized in circular components $ \sigma_\pm (\omega, q) = \sigma_{xx} (\omega, q) \pm i \sigma_{yx} (\omega,q ) .$ Using the approximation (8) for the curvature (the factor $ W(\theta, \varphi) $ is now reduced to $ W(\theta)), $ we get the following expressions for the additional terms $ \sigma_a: $
  \be 
\sigma_a(\omega, q) \approx \sigma_0 \epsilon \left [1 - i\tan \left (\frac{\pi\beta}{2}
\right) \right] (u \chi_\pm)^\beta .
          \ee
Here, $ \chi_\pm = 1 \mp \Omega/\omega + i /\omega \tau, \ \epsilon $ is a dimensionless parameter whose value is determined with the relative number of the conduction electrons concentrated at the "anomalous" effective cross-section. The term $ \sigma_a (\omega, q )$ could predominate, assuming that the parameter $ \epsilon $ takes on values of the order of unity. Otherwise, this term corresponds to the first correction to the main approximation for the conductivity. We remark that Eq.(9) gives a general expression for the contribution to the transverse conductivity originating from a nearly cylindrical effective strip at any axially symmetric FS. Unlike resembling results of the earlier work \cite{thirteen} this expression is not associated with a particular form of the energy-momentum relation near the effective line.

Now, we turn to analysis of possible effect of the FS local flattening on the electron conductivity. For certainty we assume that the flattening point is located at $ \theta = \pi/2, \ \varphi = 0.$ The contribution to $ \sigma_{xx} $ from the effective strip passing through this point can be presented as the sum of two terms. The first term has the same order of magnitude as the correction
(5) and is given by    \be 
\sigma_{axx}^{(1)} (\omega, q) = i \sigma_{xx}^0 (q) \frac{\omega}{q} \sum \limits_n
\bigg < \bigg < \frac{\chi_n}{v_a} \bigg > \bigg >,
                   \ee   where
  \bea 
 &&\bigg < \bigg < \frac{\chi_n}{v_a} \bigg > \bigg > =
\frac{8}{\pi^2 p_0^2} \int \limits_{-\infty}^0 e^\eta d \eta \int \limits_0^{\pi /2} d \theta \int \limits_0^{\pi /2} \frac{d \varphi \cos \varphi \sin \theta}
{\cos^2 \theta |K(\theta ,\varphi)|}               
    \nn \\ \nn \\ & \times &
 \left [ \frac{V_{nx} (\pi /2, \varphi) \chi_n (\pi /2,
\varphi , \eta)}{v(\pi /2, \varphi)} -
\frac{V_{nx} (\theta, \varphi)\chi_n (\theta, \varphi ,\eta)}
{v(\theta, \varphi)} \right ] .\nn \\
            \eea
 Like the quantity $ v_0 $ in the expression (5), $ v_a $ also has the dimensions of velocity. The second term, which contains the contribution to the conductivity from the locally flat region can be represented in the form
 \bea 
  &&\sigma_{axx}^{(2)} (\omega, q) = \sigma_{xx}^{(0)} (q) \frac{8}{\pi^2} u_0 \sum \limits_n V_{n0}^x
\int _{-\infty}^0 e^\eta d \eta   \nn \\ \nn \\ & \times &
\int _0^{\pi /2}\frac{\chi_{n0} (\eta) \sin \theta d
\theta}{u_0^2 \chi_{n 0}^2 (\eta ) - \cos^2 \theta}
\int _0^{\pi /2} \frac{\cos^2 \varphi d \varphi}{|K (\theta, \varphi)|}.
            \eea
Here, $ u_0 = \omega /q v (\pi /2, 0); \; V_{n0}^x =V_{nx}
(\pi /2,0); \; \chi_{n0} (\eta) = \chi_n(\pi /2, 0, \eta).$
The main contribution to the integral with respect to $ \theta $ and $ \varphi $ comes from the neighborhood of the point where the FS is flattened and $ K(\theta,\varphi)$
turns zero. When the singularity in the denominator of Eq.(12) is well pronounced (the FS is nearly flat in the vicinity of the corresponding point) this term could predominate over the "usual" first correction to the conductivity given by Eqs.(5),(10) as well as this occurs when the FS includes a nearly cylindrical effective strip.

To bring more certainty into further analysis we adopt the following energy-momentum relation:
 \be 
E{\bf (p)} = \frac{p_1^2}{2 m_1} \left (\frac{p_y^2 + p_z^2}{p_1^2} \right)^l
+ \frac{p_2^2}{m_2} f \left (\frac{p_x}{p_2} \right ).
            \ee
 The corresponding FS is a lens whose radius and thickness are $ p_1 $ and $ p_2 $, respectively;
 $ f (p_x/p_2) $ is an even function of $ p_x $
which increases monotonically for $ p_x > 0,$ provided that $ f (0) = 0, \; f (1) = 1. $ For $ l = 1 $ and $ f (p_x/p_2) = (p_x /p_2)^2, $ the energy-momentum relation (13) describes an ellipsoidal FS. In this case $ m_1 $ and $ m_2 $ are the principal values of the effective mass tensor. The lens could be a part of a multiply connected FS. For example, electron lenses are included in the FSs of cadmium and zinc. It is worth to mention that the cyclotron resonance in a normal magnetic field was observed in both metals.

The Gaussian curvature at any point of the lens corresponding to (13) is given by the expression
    \bea 
K{\bf (p)} &=& \frac{l}{m_1 v^4} \left (\frac{p_y^2 + p_z^2}{p_1^2} \right )^{l-1} \\
  & \times &
\left [(v_y^2 + v_z^2) \frac{\partial v_x}{\partial p_x} +
v_x^2 \frac{l(2l - 1)}{m_1}
\left (\frac{p_y^2 + p_z^2}{p_1^2} \right )^{l-1} \right ].
          \nn  \eea

For the $ l > 1,$ the Gaussian curvature $ K(p_x, p_y, p_z)$ vanishes at the points $ (\pm p_2, 0,0)$ coinciding with the vertices of the lens. This is illustrated in the Fig. 1.
In view of axial symmetry of the lens, the curvatures of
both principal cross-sections turn zero at these points, so the vertices are the points of flattening of the FS. As usual, we assume the magnetic field to be directed along the "z" axis, so, the flattening points at the vertices of the lens belong to an effective strip.

\begin{figure}[t] 
\begin{center}
\includegraphics[width=5cm,height=5cm]{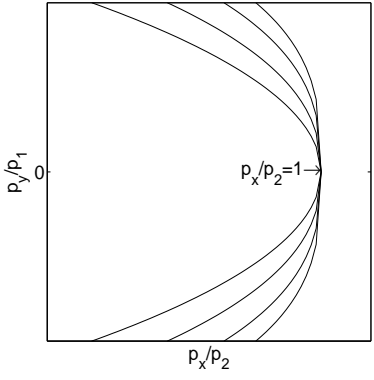}
\caption{Local flattening of the lens described with Eq.(18) near the points $ (\pm p_z,0,0) . $ The curves representing the profiles of the lens cross-sections at $ p_z = 0 $ are plotted for $ s = 0,0.2,0.4,0.5 $ (from the inner to the outer curve). } \label{rateI}
\end{center}
\end{figure}

While carrying out integration in (12), we can use an approximate expression for the curvature of the lens in the vicinity of the flattening region:
  \be 
K(\theta, \varphi) = \frac{1}{p_a^2} (\sin^2 \varphi + \cos^2 \theta)^s.
            \ee
Here, $ p_a^2 = p_1^2 (s + 1) \big [4(s + 1)^2 m_1^2 v(\pi /2,0)^2 \big / s^2 p_1^2 \big ]^{s +1} $ is a constant having the dimensions of the momentum squared, $ s = 1 - 1\big /(2l - 1); \; 0 < s < 1 $ for $ l > 1.$ The closer the value of s to unity, the more flattened the FS in the vicinity of the point where $ \theta = \pi /2, \varphi = 0.
$ For $ s \to 1, $ the FS is transformed into a plane in the
vicinity of this point. To within terms of the order of $ \sigma^{(1)}, $ the term $
\sigma_{axx}^{(2)}$ have the form:
   \bea 
 && \sigma_{xx}^{(2)} (\omega, q ) =  \sigma_{xx}^0 (q) \epsilon_s       \\ \nn \\  &\times &
\sum \limits_n
\left \{ \big [1  -  i \cot (\pi s)\big ]
\overline{(u_0 \chi_{n 0})}^{1-2s} + \frac{2 i}{\pi}
\overline{(u_0 \chi_{n0})} \right \}, \nn
            \eea
where the following notation has been used:
   \bea 
\epsilon_s &=& \frac{2}{\pi^{1/2}} \frac{p_a^2}{p_0^2}
\frac{\Gamma (s - 1/2)}{\Gamma (s)};
             \\ \nn \\
\overline{(u_0 \chi_{n0})}^\beta &=& V_{n0}^x \int_{-\infty}^0 e^\eta \big [u_0 \chi_{n 0} 
(\eta) \big ]^\beta  d  \eta;  
           \eea
 and $ \Gamma(x) $ is the gamma function.
The expansion of $\sigma_{yy} $ as well as expansions for nondiagonal conductivity components, do not contain any such term since in this case the electron velocity component $ v_y $ vanishes at the points of flattening of the Fermi surface.

Close resemblance of the results (9) and (16) gives an evidence that locally flattened and nearly cylindrical segments of the FSs affect the high-frequency conductivity in a similar way. In both cases an additional term appears in the expression for conductivity, whose frequency dependence is special. In the next Section we show that these "anomalous" terms could sensibly influence the frequency dependence of the surface impedance giving rise to various resonance features at $ \omega = \Omega. $

\section{III. surface impedance }

We calculate the surface impedance of a semi-infinite metal in the presence of a normal magnetic field. Calculations are carried out under the anomalous skin effect conditions, when the electron mean free path $ l $ is greater than the skin depth $ \delta. $ We choose these conditions for they are favorable for observations of resonance features at cyclotron frequency. We assume a specular reflection of electrons from the metal surface to predominate, which is typical at anomalous skin effect. We start our calculations within the model of axially symmetric FS, so the surface impedance tensor is diagonalized in circular components. We have:
     \be 
Z_{\pm} = \frac{8i\omega}{c^2} \int_0^{\infty}
\frac{dq}{4\pi i\omega \sigma_{\pm}/c^2 - q^2},      \ee
where $\sigma_{\pm}$ are
the circular components of the transverse conductivity.

To proceed we turn to the integration over a new variable
$t = (q\delta)^{-1}.$ Then we can divide the integration
range into two segments with different asymptotic behaviors of conductivity: 
    \bea
 Z_\pm & = & Z_1^\pm + Z_2^\pm,
             \nn\\ \nn \\
Z_1^\pm & =& -\frac{8i\omega}{c^2}\delta \int_0^{t_0}
\frac{dt}{1 - it^3\overline\sigma_\pm(t/\xi)},     
        \\ \nn \\ 
Z_2^\pm &=& -\frac{8i\omega}{c^2}\delta \int_{t_0}^\infty
\frac{dt}{1 - it^3\overline\sigma_\pm (t/\xi)},
                            \eea
Here $ \overline \sigma_\pm = \sigma_\pm / \sigma_0 , $ and $ \xi = v_m/(\omega\delta)$ is the anomaly parameter. The limit $ t_0$ in the integrals is chosen so that $u = t/\xi $ is smaller than unity for $t = t_0$. Under these conditions we can calculate $Z_1$ using for $\overline\sigma$ the expansion similar to (7):
   \be 
\sigma_\pm = \sigma_0 (q) (1 + \Lambda_1^\pm u + \Lambda_2^\pm u^2 + \dots).
  \ee

At the anomalous skin effect the surface impedance
can be expanded in the inverse powers of the
anomaly parameter. The main terms in this expansion originate from the addend $ Z_1.$ These terms can be
readily found by expanding the integrand in Eq.(20) in powers of the parameter $\lambda $ given by:
 \be 
\overline\sigma = 1 + \lambda.
                                 \ee
This parameter is small for $\xi>>1$. We obtain:
   \bea 
 &&Z_1^\pm = \frac{8\omega}{c^2}\delta \left
[\frac{\pi}{3\sqrt 3}(1 - i\sqrt 3) +
\frac{2 \pi}{9\sqrt 3}\frac{\Lambda_1^\pm}{\xi}
(1 + i\sqrt 3) \right.
 \nn \\ \nn \\
 & + & \left.
\frac{\Lambda_1^{\pm 2}- \Lambda_2^\pm}{\xi^2}
\left (\ln t_0 + \frac{i\pi}{3} \right ) - \frac{1}{2t_0^2} +
\frac{\Lambda_2^\pm}{3\xi^2} - \frac{\Lambda_1^\pm}{t_0\xi}
\right ] + \delta Z_1. \nn \\
                         \eea

To arrive at the result (24) we used three first terms in
the expansion of $ \overline \sigma $ in powers of the small parameter $ u $ and we kept the terms of the order of $ (t/\xi)^2 $ in the expression for $ \lambda . $ Taking into account next terms in the expansion of $ \overline \sigma $ in powers of $ u $ or keeping next terms in the expansion of $ \lambda $ we obtain that corresponding integrals diverge. Therefore we cannot expand the correction $ \delta Z_1$ in the inverse powers of the anomaly parameter and we leave it in the form:
 \be 
\delta Z_1^\pm = \frac{8\omega}{c^2}\delta \int_0^{t_0}\!\!\!
\left(\frac {1}{t^3\overline\sigma_\pm(t/\xi)} - \frac {1}{t^3} +
\frac{\Lambda_1^\pm}{\xi t^2} -
\frac{\Lambda_1^{\pm 2} - \Lambda_2^\pm}{\xi^2 t}\right) dt.
                                      \ee
This expression is correct to the terms of the order of
$(\xi)^{-3}$.

When we calculate the second term in the expression for the
impedance, we can ignore the unity in the denominator of the
integrand in Eq.(21):
       \be 
Z_2^\pm = \frac{8\omega}{c^2}\delta \int_{t_0}^\infty
\frac {dt}{t^3\overline\sigma_\pm (t/\xi)} .
                                     \ee
Eqs.(24)--(26) include the parameter $t_0$ which has been
introduced arbitrarily. As should be expected, however, we have $dZ/dt_0 = 0$. So, the impedance does not depedend of $t_0$. Therefore, we can select a specific value of the parameter $t_0$ (assuming that $t_0 \sim\xi $) and then derive various equivalent forms of Eqs.(24)--(26).

Now, we perform analytic continuation of the integrands in Eqs.(25) and (26) to the first quadrant of the complex plane. To calculate the term $ \delta Z_1$ we use a path shown in the Fig. 2a. The path includes segments of real and imaginary axes and a circular arc of the radius $ t_0. $ To calculate $ Z_2 $ we choose the path shown in the Fig. 2b. 

  \begin{figure}[t] 
\begin{center}
\includegraphics[width=8.8cm,height=4.4cm]{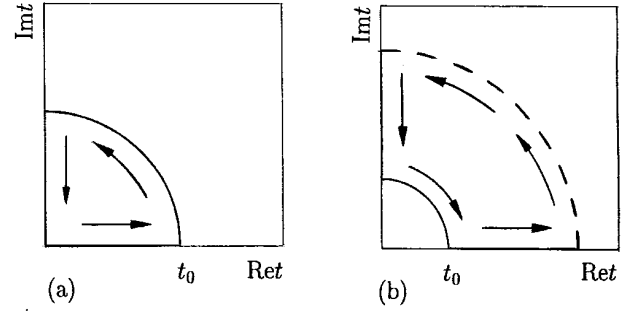}
\caption{Contours used for calculation of the quantities (a) $ \ \delta Z_1\ $ and (b) $\ Z_2.$ } \label{rateI}
\end{center}
\end{figure}

We put $ t_0 = \xi , $ and after some straightforward calculations we arrive at the result:
   \bea 
Z^\pm &=& \frac{8\omega}{c^2}\delta
\left [\frac{\pi}{3\sqrt 3}(1 - i\sqrt 3) +
\frac{2 \pi}{9\sqrt 3} \frac{\Lambda_1^\pm}{\xi} (i - \sqrt 3)
 \right.  \nn \\ \nn \\
&+ &\frac{\Lambda_1^{\pm 2} - \Lambda_2^\pm}{\xi^2}
\left (\ln \xi+ \frac{i\pi}{3} \right)    + \frac{1}{2 \xi} + \frac{\Lambda_2^\pm}{3\xi^2}
  \nn \\ \nn \\
 &-&  \left.\frac{\Lambda_1^{\pm 2}}{2\xi^2} -
\frac{i\Lambda_1^\pm}{ \xi^2} -
Y_1^\pm - Y_2^\pm \right ] .
                                        \eea
Here we have
         \bea 
Y_1^\pm &= & \frac{1}{\xi^2} \int_{1}^{\infty}
\left(\frac{y}{\overline\sigma_\pm(i/y)} - y + i\Lambda_1^\pm +
\frac{\Lambda_1^{\pm 2} - \Lambda_2^\pm}{y} \right) dy;    
                                 \nn\\ \\
Y_2^\pm & = & \frac{1}{\xi^2} \int_{0} ^1
\frac{y}{\overline\sigma_\pm (i/y)} dy.
                                        \eea

The coefficients $ \Lambda_{1.2}^\pm $ equal:
  \be 
\Lambda_1^\pm = - \frac{ig}{\pi} b \chi_\pm; \qquad
\Lambda_2^\pm = d \chi_\pm^2
   \ee
where the values of the dimensionless constants $g,b $ and $ d $ are determined with the FS shape. For a spherical FS our result (27) agrees with the similar result derived earlier \cite{one}.

 The result does not reveal resonance at the cyclotron frequency, since not a single term in the expression (27) for $ Z_- $ exhibits a resonance behavior at $ \omega = \Omega. $ This proves that the cyclotron resonance in a normal magnetic fields cannot occur in metals whose FSs possess finite and nonzero curvature, and emphasizes the part played by anomalies in the FS curvature in the appearance of the effect. We analyze the latter below.

We first consider the case when the effective cross-sections of the FS include one of a zero curvature. It can be the only nearly cylindrical cross-section if it is the central one.
Otherwise, the effective cross-sections are combined in pairs which are symmetrically arranged with respect to a plane $p_z = 0$. The contribution of the nearly cylindrical cross-sections to the surface impedance depends on the relative number of the effective electrons associated with them. When a considerable part of conduction electrons is concentrated at the nearly cylindrical effective segments of the FS $ (\epsilon \sim 1 ) ,$ then the anomalous contribution to the conductivity $ \sigma_a $ is the principal term in the expansion of the conductivity in powers of $u, $
and it strongly contributes to the surface impedance. As a result the principal term of the impedance in the case of anomalous skin effect is given by

 \begin{figure}[t] 
\begin{center}
\includegraphics[width=8.0cm,height=5.8cm]{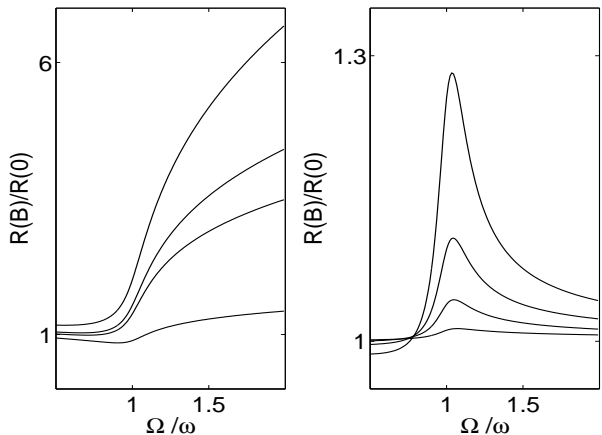}
\caption{ Magnetic field dependence of the real part of the surface impedance near the cyclotron resonance assuming that the FS is axially symmetric at an effective cross-section for $\epsilon = 0.8, $ (left panel) and $ \epsilon = 0.01 $ (right panel). The curves are plotted for $ \omega \tau = 10, \ \xi^3 = 10^4, \ \beta = - 0.8, -0.7, - 0.6, -0.4 $ (from the top to the bottom).} \label{rateI}
\end{center}
\end{figure}
                   \bea 
  Z_\pm &=& \frac{8 \pi \omega}{c^2} \delta 
\left(\frac{\xi\beta}  {\epsilon\cos(\pi\beta/2)} 
\right)^{1/(\beta + 3)}
\frac{\cot(\pi/(\beta + 3))}{\beta + 3}
          \nn \\ \nn \\  &\times &
\left [1 - i\tan \left(\frac{\pi}{\beta + 3}\right)
\right ](\chi_\pm)^{-\beta /(\beta + 3)}.
                             \eea 
  This expression shows a very special dependence of the surface impedance on the magnetic field. Near the cyclotron frequency the real and imaginary parts of the component $ Z_- $ swiftly rise. The increase in $ R (B) $ is no smaller in order than $ R (0) $ at the same frequency. In strong magnetic fields $ (\omega << \Omega) $ the value of $ R ( B) $ rises proportionally to $ ( \Omega/\omega )^{-\beta/(\beta+3 )}$ as shown in the Fig. 3 (see the left panel). When the parameter $ \beta $ takes on values close to zero, $ R(B)$ almost levels off at strong magnetic fields whereas in the limit $ \beta \to -1 \ \ R(B) \sim \sqrt{\Omega/\omega}. $

No experimental observations of such impedance behavior have been reported. The reason is that we hardly can expect the parameter $ \epsilon $ to take on values of the order of unity in real metals. We must rather expect that the relative number of electrons associated with the nearly cylindrical cross-section is small
$ \epsilon <<1 $. Then the principal term in the
surface impedance expansion in inverse powers of the anomaly parameter remains the same, as given by Eq. (27). However, we obtain additional higher-order terms corresponding to the contribution of the anomalous effective cross-section. The
main approximation for the correction originating from a
nearly cylindrical effective strip equals:    
  \bea 
 Z_a^\pm & = &  - \frac{8 \pi v_F}{9c^2 \xi}
\left (\frac{\chi_\pm}{\xi} \right )^\beta \epsilon
\left (\frac{\beta + 1} {\cos(\pi\beta/2)} \right )  
 \\  \nn \\ &\times &
\cot \left (\frac{\pi(\beta + 1)}{3} \right )
\left [1 - i\tan \left (\frac{\pi(\beta + 1)}{3} \right )\right ]. \nn
           \eea
  A comparison of Eqs.(32) and (27) shows this the term $Z_a$ is the largest among all corrections added to the principal term in the surface impedance approximation. This correction describes resonance features at the cyclotron frequency in both real and imaginary parts of the impedance component $ Z_- $ (Fig. 3, right panel). The shape of the peak in the real part of
the impedance resembles that recorded on potassium \cite{two}. The peak height depends on the value of the parameter $ \epsilon. $ For $ \epsilon \sim 10^{-2},
\omega\tau
\sim 10$, and $\xi^3\sim 10^4$ the resonance amplitude in
Re$ Z$ is approximately $10^{-2}$ of the principal term which agrees with the experiments of \cite{two} on potassium, as well as with experiments on organic metals \cite{eleven,twelve}.

In the previous section we showed that the "anomalous" contribution to the conductivity also could appear when an effective strip on the FS includes a locally flattened segment (see (16)). The occurence of such term in the conductivity can influence the surface impedance and cause resonance at the cyclotron frequency. Assuming the electrons to be specularly reflected we can write the following expressions for the surface impedance components:  \be 
Z_{\alpha \beta} = \frac{8 i \omega}{c^2} \int_0^\infty \left (\frac{4 \pi i \omega}{c^2} \sigma - q^2 I \right )_{\alpha \beta}^{-1} d q.
            \ee
 Here $ I_{\alpha \beta} = \delta_{\alpha \beta}.$
Substituting the obtained expression for the conductivity components into (33) we can evaluate the first terms in the expansion of $ Z_{xx} $ in the inverse powers of the anomaly parameter $ \xi: $
 \bea 
  & & Z_{xx} = \frac{8 \omega \delta}{c^2} \left \{
\frac{\pi}{3 \sqrt 3}(1 - i \sqrt 3) - \frac{2\pi}{9 \sqrt 3}
\frac{v_m}{\xi} ( 1 + i \sqrt 3) \right.
    \nn \\ \nn \\ & \times &
\sum \limits_n \left (
\bigg < \frac{\chi_n}{v_0} \bigg > +
\bigg < \bigg < \frac{\chi_n}{v_a} \bigg > \bigg > -
\epsilon_s B_s \sum \limits_n \left (\frac{\overline{\chi_{n 0}}}{\xi} \right )^{1-2s} \right \}. \nn \\
            \eea
Here the complex constant $ B_s $ is defined as
 \bea 
 B_s & = &  \frac{2(1 - s)}{\sqrt 3} \left (\frac{v_m}{|v(\pi /2; 0)|} \right )^{1 - 2s}
         \\ \nn \\ & \times &
\frac{\cot \big [2\pi /3(1 - s) \big ]}{\sin (\pi s)}
\left [1 - i \tan \left (\frac{2 \pi} {3} (1 - s) \right ) \right ]. \nn
            \eea

The results show that flattening points on an effective strip of the FS could bring an additional term in the expression for impedance, in the same way as in the case when one of the effective strips on th FS is nearly cylindrical. The emergence of this contribution to the surface impedance (the last term in (34)) is associated with a broadening of the strip of effective electrons due to the presence of flattening points on it.

The extent to which the locally flat regions of the FS
affect the surface impedance is determined by the relative number of the effective electrons associated with the flattened regions and the parameter $ s $ characterizing the
degree of flattening of the FS in the vicinity of the flattening point. If $ s $ considerably differs from zero and $ \epsilon_s $ is not too small, this contribution may turn out to be significant and lead to the appearence of noticeable resonance type singularities in the frequency/field
dependencies of the surface impedance.When $ s >
1/2, $ the resonance peak could be observed in the impedance itself, while for a less pronounced flattening of the FS $(0 < s < 1/2) $ the resonance singularity is manifested only in the field and frequency dependencies of the derivative of impedance. The experimental results obtained in Refs. \cite{three,four} for Cd and Zn indicate that in all probability, the latter of the above possibilities is
realized in these metals. In other words, the flattening of the FS at vertices of the electron lens is moderate $ (0 < s < 1/2).$ However, the first possibility $ ( s > 1/2)$ cannot be ruled out either, since a strong flattening of the FS could be accompanied with a relatively small value of the parameter $ \epsilon.$ In this case, the
resonance singularity of the surface impedance may not be revealed whereas the stronger singularity in the derivative of the impedance for the same $ \epsilon$ is
exhibited quite clearly. The derivative of the real part of the contribution to the impedance from a locally flat region has the form:
   \be 
\frac{d R}{d B} = \frac{\epsilon a_s}{\xi^{2(1-s)}}Y_s (\chi),
            \ee
      where
 \bea 
  a_s &=& \frac{16}{\sqrt 3}
\frac{\delta |e|}{(\pi /2)c^3} V_{10}^x
\left (\frac{v_m}{v(\pi /2,0)} \right )^{1-2s}
      \nn \\ \nn \\  & \times &
\frac{1 - s}{\sin(\pi s)\sin \big [2\pi (1 - s) \big / 3 \big ]},
            \eea
and the function $ Y_s (\chi),$ which describes the shape of the resonance curve, is given by:
 \bea 
Y_s (\chi) & = & |\chi|^{-2s} \cos \left \{2s \arctan
\left (\frac{\chi''}{\chi'} \right ) \right.
   \nn \\  \nn \\ & + & \left.
  \frac{2 \pi}{3} (1 - s) +
\pi s \theta (\Omega - \omega) \right \}.
            \eea
 Here,
  \bea 
 \theta (x)& =& \left \{ 
\begin{array}{l} 0\qquad x \le 0, \\ 1\qquad x > 0,
 \end{array} \right.
     \nn \\ \nn \\
\chi & = & \chi' + i \chi'' = 1 -
\frac{\Omega (\pi /2; 0)}{\omega}   \nn \\ \nn \\
  &+& i \left (\frac{1}{\omega \tau} -
\frac{\Delta (\pi /2; 0)}{\omega} \overline \eta \right ).            \eea

While deriving this formula (36), we have taken into account only the term with $ n = 1 $ in the sum over $ n $ in the expression for the resonance term in (34), since this result describes the field dependence of $ d R /d B $ in the region of magnetic field corresponding to the closeness of $ \Omega $ and $ \omega.$ The contribution from the harmonics of cyclotron resonance can be taken into account by considering other terms in the sum over $ n $ in (34). Finally, the quantity $ \overline \eta $ in the expression for $ \chi $ is defined as follows:
 \be 
\chi (\overline \eta) = \int_{-\infty}^0 e^\eta \chi (\eta) d \eta.
            \ee

The resonance nature of the dependence of $ dR / dB $ is manifested under the condition of smallness of the imaginary part of $ \chi. $ Apart from the obvious condition $ \omega \tau >> 1, $ the inequality $ |\Delta (\pi /2, 0)\overline \eta | << \omega $ must also be satisfied in order to keep the value of the above quantity low. If the FS
flattening points are located at the vertices of an electron lens, the effective electron velocity vector slightly varies during its motion along the part of the cyclotron orbit passing through the flat region of the lens. Therefore,
the quantity $ \Delta (\pi / 2,0)$ can be expected to be small here. However, if the magnetic field is tilted away
so that the point with coordinates $ \theta = \pi /2, \; \varphi = 0$ on the Fermi surface no longer coincides with the flattening point, the value of $ \Delta (\pi /2, 0) $ may increase considerably, and the resonance can no longer be revealed. This is one of the reasons behind the weakening of the resonance as the magnetic field deviates from the direction corresponding to the maximum amplitude of the resonance observed in the experiments of \cite{three}. Similar effect is well known for quasi two dimensional metals as discussed in \cite{thirtyfour,thirtyfive,thirtysix}.

\begin{figure}[t] 
\begin{center}
\includegraphics[width=8cm,height=5.8cm]{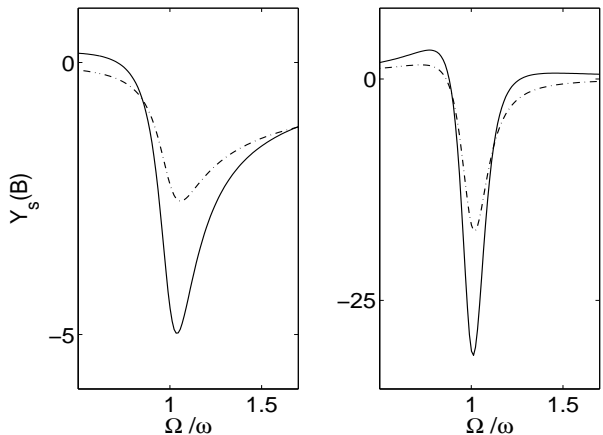}
\caption{ Plots of the function $ Y_s (\chi) $ versus magnetic field for moderate (left panel) and strong (right panel) flattening of the FS near the vertices of the lens (18). Curves are plotted assuming that $ \Delta (\pi/2,0) = 0, \ \omega \tau = 10 $ for $ s = -0.25,-0.375 $ (left panel, dashed line and solid line, respectively) and $ s = -0.625,-0.75$ (right panel, dashed line and solid line, respectively).} \label{rateI}
\end{center}
\end{figure}

Figure 4 displays the field dependence of the function $ Y_s (\chi) $ for certain values of the parameter $ s.$ A comparison with Fig. 5, which contains the recording for $ dR/dB $ obtained in Refs. \cite{three,four}
shows that the shape of the resonance lines described by (36) is in fairly good agreement with the experimental results. The variation of the amplitude and shape of the resonance lines with decreasing $ s $ resembles the variation of the experimental resonance lines upon an increase in the angle $\Phi$ between the magnetic field and the axis $ [11\overline 20]. $ The latter rests in the plane perpendicular to the symmetry axis of the electron lens included in the FSs of cadmium and zinc. The above similarity is due to the fact that as the magnetic field deviates from the axis $ [11\overline 20],$ the effective cross-section of the electron lens no longer passes through the flattening points at the vertices of the lens but still remains quite close to them for small angles of deviation. So, it can be assumed that if the angle $ \Phi $
does not exceed a certain critical value $ \Phi_0 $ characterizing the size of the flattening region on the FS, the parameter $s $ decreases with increasing $ \Phi,$ but remains nonzero. The value $ s = 0 $ corresponds to angles $ \Phi \ge \Phi_0, $ when the cyclotron orbit of effective electrons no longer passes through the locally flat region on the FS. 

  \begin{figure}[t] 
\begin{center}
\includegraphics[width=8.8cm,height=6.8cm]{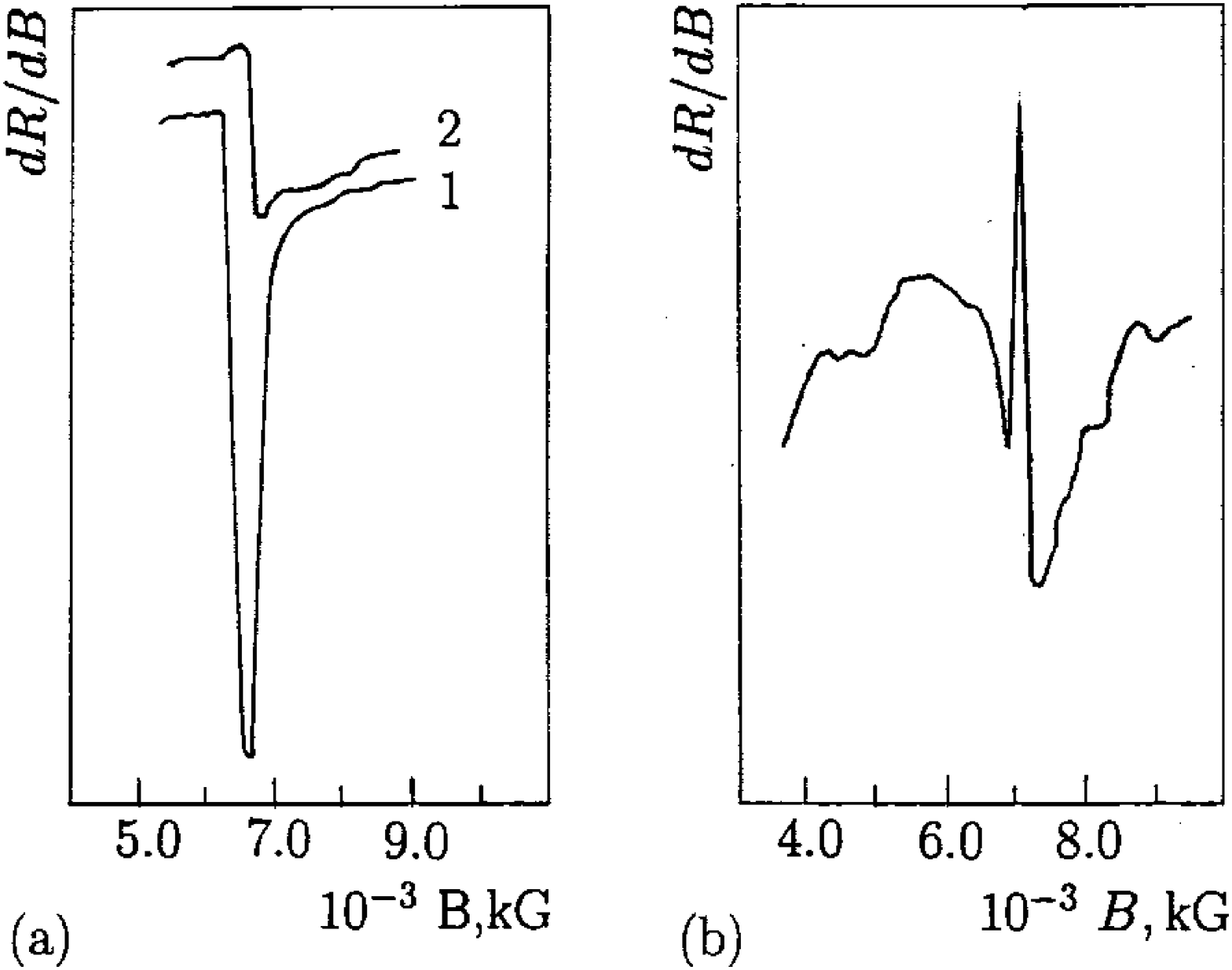}
\caption{
Resonance features in the magnetic field dependence of $ dR/dB $ recorded in experiments of \cite{three,four} on cadmium (a) and zinc (b). Curve 1 is plotted for $ \bf B || [11\overline20]$ whereas curve 2 corresponds to the case when the angle between $ \bf B $ and $[11\overline20] $
equal $8^o $ (a). Amplification at the recording of the
curve 2 compared to that of the curve 1 is $ 10:1 $.
} \label{rateI}
\end{center}
\end{figure}

\section{IV. Discussion}

It is known that the local geometry of the FS could affect high-frequency properties of metals. When the effective belts on the FS are nearly cylindrical in shape or include flattened segments, this brings changes in observables. These changes could be revealed for fitting directions of propagation of an external disturbance. As a result, specific angular dependences are to be manifested in the observables. The effect of the local geometry on frequency and angular dependences of the dispersion and attenuation of ultrasound waves, and the surface impedance of a metal was predicted and analyzed before within some simple models of the FS \cite{eighteen,nineteen,tventy,tventyone,tventytwo,tventythree,tventyfour,tventyfive,tventysix}. The present analysis shows that local geometry of the FS also could give rise to the cyclotron resonance in the normal magnetic field in good metals.

A simple explanation of the effect is that the contributions from flattened or nearly cylindrical segments of the FS to the electron density of states could be sensibly greater than those from the rest of the surface. Therefore the number of effective electrons moving inside the skin layer is increased which provides favorable conditions to the resonance to be manifested.

We can show that even slight distortion of the original Fermi sphere caused by the crystalline fields brings along curvature anomalies. Adopting the nearly-free-electrons approach we arrive at the energy-momentum relation for conduction electrons:
     \be 
E= \frac{{\bf k}^2}{2 m} +
\frac{{\bf g}^2}{2 m} - \frac{1}{m}
\sqrt{({\bf k \cdot g})^2 + m^2 V^2} ,
                          \ee
where $ m$ is the effective mass; $ {\bf k = g - p;
\; g} = \hbar {\bf G}/2; \; \bf G $ is a reciprocal lattice wave vector; $ V $ is the Fourier component of the potential energy of electron in the lattice field which corresponds to the vector $ \bf G.$ Within the nearly-free-electron model the energy $ V $ is assumed to be small compared to the Fermi energy of electrons $ E_F, $ so we introduce a small parameter $ \zeta = \sqrt{V/E_F}. $

The corresponding FS represents,
roughly speaking, a sphere with ''knobs'' located at those segments which are close to the boundaries of the Brillouine zone. Inflection lines of zero curvature pass along the boundary between knobs and the main body of the FS. A FS segment including a knob and its vicinity is axially symmetric, and the symmetry axis is directed along the corresponding reciprocal lattice vector. In further analysis we single out such FS segment to consider it separately. For certainty we choose the coordinate system whose $"x"$ axis is directed along the reciprocal lattice vector. Within the chosen segment the FS curvature is described with the expression:

     \be 
  K = \frac{m^2 v_x^2 + p_\perp^2 d v_x/ dp_x}{(p_\perp^2 + m^2 v_x^2)^2}
    \ee
  Where $ p_\perp^2 = p_y^2 + p_z^2. $

Equating the FS curvature to zero, and using the energy-momentum relation (41) we find the values $ p_{x0} $ and $ p_{\perp 0} $ corresponding to the inflection line. We get:
  \be 
p_{x0} \approx p_F \left ( 1 - \frac{\zeta^2}{\sqrt2} \right); \qquad p_{\perp 0} \approx p_F \zeta;
 \ee
 where $ p_F $ is the radius of the original Fermi sphere. Now, we can expand the variable $ p_x $ in powers of $ (p_\perp - p_{\perp 0} )$ near the zero curvature line. Taking into account that $ d^2 p_x / d p_\perp^2 $ turns zero at points belonging to the inflection line, and keeping the lowest-order terms in the expansion, we obtain:
 \be 
p_x \approx p_{x 0} - \frac{\zeta}{\sqrt2} (p_\perp - p_{\perp 0}) - \frac{p_F}{\sqrt2 \zeta}
\left ( \frac{p_\perp - p_{\perp 0}}{p_F} \right )^3.
 \ee
Substituting this approximation into Eq.(41) we arrive at the following energy-momentum relation:  \be 
E ({\bf p}) = \frac{p_x^2}{2m} + \frac{2}{\zeta} \frac{p_F^2}{2m}
\left ( \frac{p_\perp - p_{\perp 0}}{p_F} \right )^3 .
 \ee
The latter could be employed near zero curvature line where $ p_{\perp 0} << p_F. $ Omitting $ p_{\perp 0} $ we arrive at the energy-momentum relation of the form (13) where $ l = 3/2 \ \ (s = 0.5). $

This semiquantitative analysis shows that the form of the energy-momentum relation (13) could be reasonably justified within a nearly free electron approximation. In real metals the flattening of some FS segments could be manifested even more.
Inflection lines as well as points of flattening exist on the FSs of the most of the metals, therefore we can expect cyclotron resonance in a normal magnetic field to be manifested there for suitable magnetic field directions.
There is an experimental evidence that "necks" connecting quasispherical pieces of the FS of copper include nearly cylindrical belts \cite{tventyseven}. When the magnetic field is directed along the axis of a "neck" (for instance, along the [111] direction in the quasimomenta space), the extremal cross section of the "neck" could be expected to run along the nearly cylindrical strip where the FS curvature turns zero. It is also likely that the FS of gold possesses the same geometrical features for it closely resembles that of copper. As was already mentioned the cyclotron resonance in the normal field was observed in organic metals of the $(\alpha-$BEDT--TTF$)_2 $MHg(SCN$)_4 $ group and some other layered conductors. The FSs of these materials are supposed to be a set of weakly rippled cylinders (isolated or connected by links) \cite{tventyeight,tventynine,thirty,thirtyone,thirtytwo,thirtythree,thirtyfour,thirtyfive,thirtysix} whose axes are perpendicular to the layers. Usually, the profiles of the cylinders are described adopting the tight-binding approximation for the electrons energy-momentum relation. Within this model the FS curvature is finite and nonzero at the effective cross-sections assuming that the magnetic field is directed along the axes of the cylinders. However, experimental results of \cite{eleven,twelve} do not agree with this model very well. Our approach was succesfully used
to analyze the effects of various profiles of the cylinders on high-frequency properties of the layered conductors. Also it was successfully applied to describe special features of quantum oscillations in the elastic constants in these materials \cite{thirtyseven}, as well as in conventional metals \cite{thirtyeight}.

Finally, this work presents novel results concerning the effect of local anomalies of the FS curvature in the initiation of the cyclotron resonance in conventional metals in a normal magnetic field. We show that this phenomenon could occur in 3D metals provided that their FSs include nearly cylindrical strips or flattened segments, and we give a theoretical explanation to the existing experimental evidence of the works \cite{two,three,four}. The presented results resemble those obtained for quasi-2D organic metals which proves that the considered effect has the same nature and origin in both kinds of materials.
The results are based on thorough theoretical analysis and give further understanding of high-frequency properties of metals.

\vspace{2mm}

{\it Acknowledgments:}
We thank G.M. Zimbovsky for help with the manuscript. This work was supported in part by NSF Advance program SBE-0123654.



\end{document}